\documentclass[5p,twocolumn,times,number]{elsarticle}
\usepackage{graphicx}
\usepackage{amsmath}
\usepackage{lineno}

\begin{document}
\begin{frontmatter}

\title{Development of Multi-gap Resistive Plate Chamber (MRPC) for Medical~Imaging}
\author[label1]{A.~Banerjee}
\author[label1]{A.~Roy}
\author[label1,label2]{S.~Biswas\corref{cor}}
\ead{saikat.ino@gmail.com}
\author[label1]{S.~Chattopadhyay}
\author[label1]{G.~Das}
\author[label1]{S.~Pal}

\cortext[cor]{Corresponding author}

\address[label1]{Variable Energy Cyclotron Centre, 1/AF Bidhan Nagar, Kolkata-700 064, India}
\address[label2]{GSI Helmholtzzentrum f\"{u}r Schwerionenforschung GmbH, Darmstadt, Germany-64291}

\begin{abstract}
The low cost and high resolution Multi-gap Resistive Plate Chamber (MRPC) opens up a new possibility to find an efficient alternative
detector for the Time of Flight (TOF) based Positron Emission Tomography, where the sensitivity of the system depends largely on the
time resolution of the detector. In a layered structure, suitable converters can be used to increase the photon detection efficiency.
In this paper results of the cosmic ray test of a four-gap bakelite-based prototype MRPC operated in streamer mode and six-gap glass-based MRPC operated in avalanche mode
 are discussed.

\end{abstract}
\end{frontmatter}

\section{Introduction}
\label{}
Conventional PET scanners use expensive scintillators for detecting photon pairs with very high efficiency and excellent energy resolution. However,
the process of detecting scintillation light using PMT and associated electronics for charge integration circuits results in a large dead time.
In addition to that, the scanners suffer from the limitations of a short Field of View (FOV) [15-25 cm] and relatively poorer position resolution.
In a Time-of-Flight (TOF) PET scanner, detectors with very good time resolution when positioned around the object can be used to record the time of
 arrival of two photons. The time difference between two photons can be used to locate the point of annihilation. TOF information of the photons when coupled to the conventional PETs, improve the image quality significantly.

Apart from the limitations of the scintillator-based systems mentioned earlier, high quality scintillators are very expensive and efforts are being
made to look for not-so-expensive alternatives. Extremely good time (50 ps) and position ($\sim$ mm) resolution make the relatively less expensive
MRPC-based TOF-PETs to be good candidates \cite{EC374,AB508,CL602,PF}. In MRPC-based system, the location of the origin of photons can be obtained in 3-dimensions. The detector
cell provides the position in transverse plane (x-y plane) and the time of arrival on the detector gives the distance in z-direction.

An R\&D effort on the feasibility study of MRPC in PET imaging has been initiated at VECC. Small size prototype MRPCs have been built with both bakelite and glass and tested with cosmic rays in streamer and avalanche mode respectively. In this article, some experimental results of a four-gap (gas gap: 0.6 mm $\times$ 4) prototype bakelite MRPC, operated in streamer mode with argon, isobutane and tetrafluroethane (R-134a) in the volume ratio of 55/7.5/37.5 and a six-gap (gas gap: 0.2 mm $\times$ 6) prototype glass MRPC operated in avalanche mode with R-134a/isobutane/ SF$_6$ in 95/4.5/0.5 are reported.

\section{Test setup and method of calculation}
\label{}
The MRPCs were tested in the same cosmic ray test bench described in Ref.~\cite{SB109}. Three scintillators, two placed above the MRPC and one below, were been used to construct the cosmic ray telescope. To measure the timing properties of the detectors the triple coincidence of the signals obtained from the three scintillators was taken as the START signal (master trigger) for the Time to Digital Converter (TDC). The STOP signal was taken from a single MRPC strip. The details of method of calculation are described in Ref.~\cite{SB110}.

\section{Results}
\label{}
\begin{figure}[b!]
\centering
\includegraphics[scale=0.28]{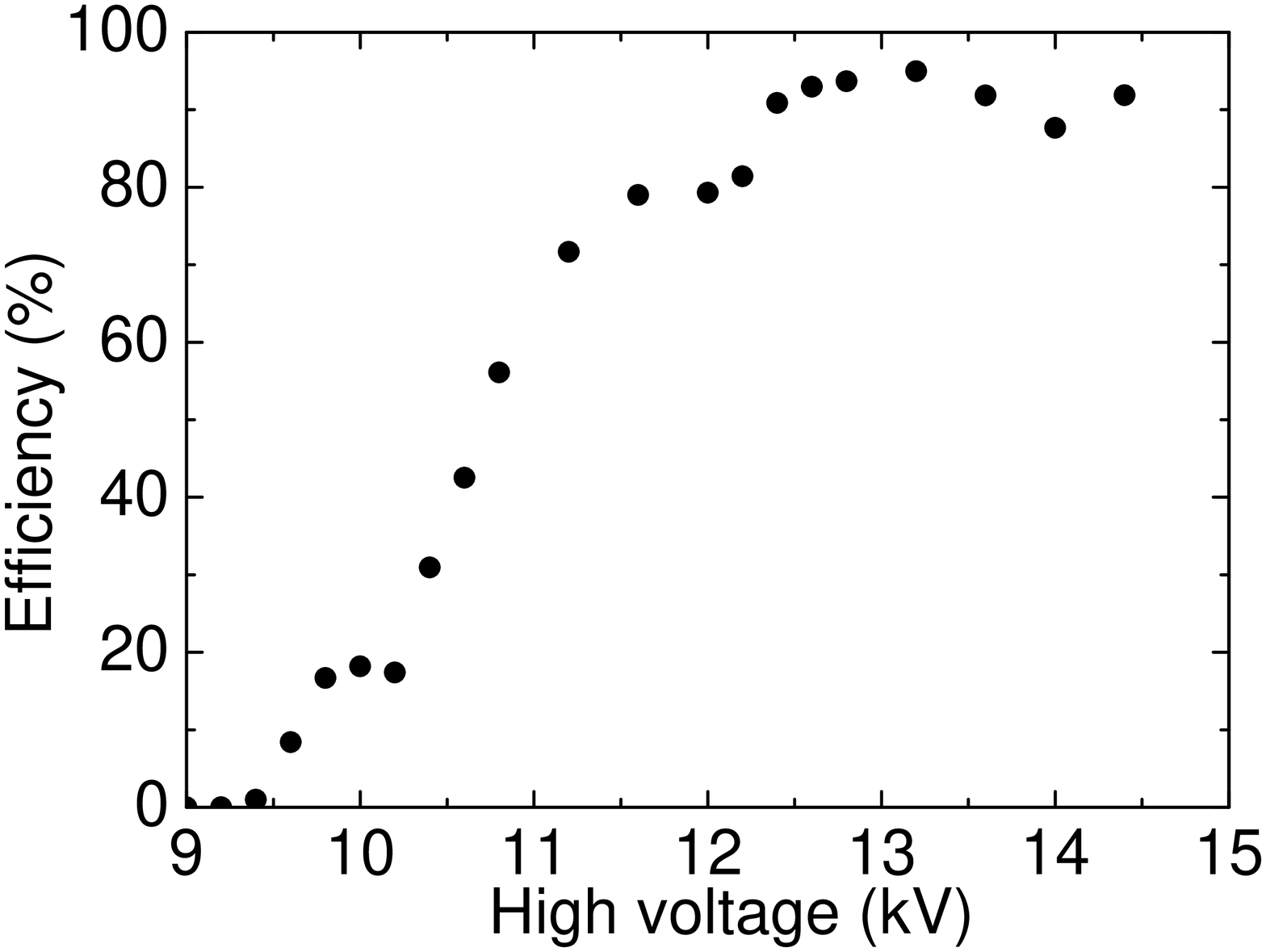}\\
\caption{The efficiency as a function of high voltage for the four-gap bakelite MRPC.  }\label{fig:1}
\end{figure}
The variation of efficiency with that of the applied high voltage (HV) is shown in Fig.~\ref{fig:1} for four-gap bakelite MRPC. The efficiency gradually increases and reaches the plateau at $\sim$95\%.

\begin{figure}[h!]
\centering
\includegraphics[scale=0.29]{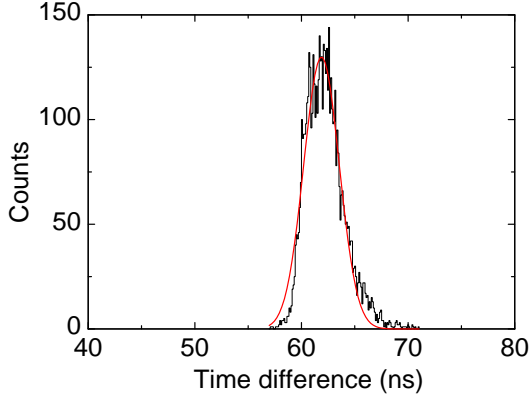}\\
\caption{The distribution of the time difference between the RPC and the master trigger for the four-gap bakelite MRPC.}\label{fig:2}
\end{figure}

Fig.~\ref{fig:2} shows the distribution of the time difference between the master trigger and the signal from one strip of the four gap bakelite MRPC. The intrinsic time resolution ($\sigma$) of the MRPC at 13.5 kV operating voltage appeared to be  : 849 $\pm$ 17 ps.
\begin{figure}[h!]
\centering
\includegraphics[scale=0.29]{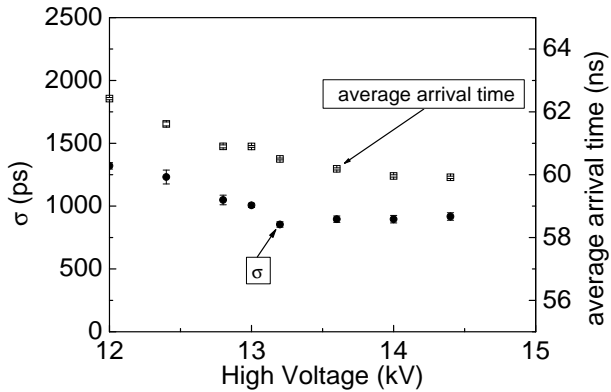}\\
\caption{The time resolution ($\sigma$) and the average signal arrival time with respect to the master trigger as a function of HV for the four-gap bakelite MRPC.}\label{fig:3}
\end{figure}
The average signal arrival time, taken as the mean position of the fitted Gaussian peak in the timing spectrum (such as in Fig. ~\ref{fig:2}) and the time resolution ($\sigma$) of the bakelite MRPC as a function of the applied HV are shown in Fig.~\ref{fig:3}. The time resolution improves and the average signal arrival time decreases with the increase of HV which is common to any gas filled detector. 

\begin{figure}[h!]
\centering
\includegraphics[scale=0.29]{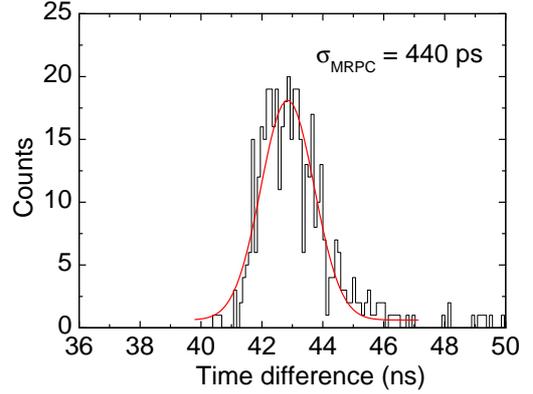}\\
\caption{The time spectrum for the six-gap glass MRPC .}\label{fig:4}
\end{figure}

The timing spectrum for the six-gap glass MRPC is shown in Fig.~\ref{fig:4} where the intrinsic time
resolution comes out to be $\sim$~440~ps ($\sigma$).
\label{}
\begin{figure}[h!]
\centering
\includegraphics[scale=0.35]{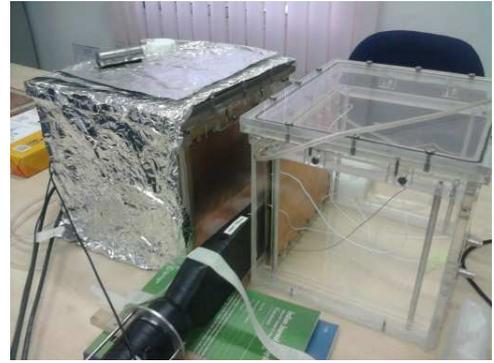}\\
\caption{The setup for glass-based MRPC system to be tested with gamma source }\label{fig:5}
\end{figure}

\section{Conclusions and outlook}
\label{}
MRPC-based PET system is potentially an attractive alternative to the scintillator-based system. As an effort to obtain a least expensive PET
imaging system capable of filtering the background, we have built a 4-gap Bakelite-based MRPC giving time resolution of $\sim$ 900 ps ($\sigma$) and
a six-gap glass-based MRPC giving timing resolution of $\sim$~440~ps~($\sigma$). Decreasing the individual gas gap and increasing the number of gaps the time resolution can be improved. Simulation to optimise the number of gaps and width of individual gap is in progress.
For the proper application in PET imaging such a system with relatively lower time resolution can be useful. This also provides a platform to build a full glass-based high resolution MRPC system for PET imaging. The test set-up for one such glass-based system is shown in Fig.~\ref{fig:5} where a Na$^{22}$ source, emitting two back-to-back 511 keV gamma ray, will be used to make the situation similar to PET imaging. The tests are in progress and the results of its performance will be communicated in a later stage.

\section{Acknowledgement}
\label{}
This work is supported by the DAE-SRC award to S. Chattopadhyay under the scheme no.2008/21/07-BRNS/2738.



\begin{thebibliography}{10}
\bibitem{EC374} E. Cerron Zeballos, et al., Nucl. Instr. and Meth. A 374 (1996) 132.
\bibitem{AB508} A. Blanco, et al., Nucl. Instr. and Meth. A 508 (2003) 88.
\bibitem{CL602} C. Lippmann, et al., Nucl. Instr. and Meth. A 602 (2009) 735.
\bibitem{PF}P. Fonte et al., Nucl. Instr. and Meth. A 449 (2000) 295.
\bibitem{SB109} S.Biswas, et al., Nucl. Instr. and Meth. A 602 (2009) 749.
\bibitem{SB110} S.Biswas, et al., Nucl. Instr. and Meth. A 617 (2010) 138.

\end{thebibliography}
\end{document}